# Behavior of hydrophobic ionic liquids as liquid membranes on phenol removal: Experimental study and optimization

**Authors and Affiliation**

Y. S. Ng [a], N. S. Jayakumar [a], M. A. Hashim [a]*

[a] Department of Chemical Engineering, University of Malaya, 50603 Kuala Lumpur, Malaysia

*Corresponding author*

| | |
|---|---|
| Email | : alihashim@um.edu.my |
| Telephone number | : +603-79675296 |
| Fax number | : +603-79675319 |
| Postal Address | : Department of Chemical Engineering, Faculty of Engineering, University of Malaya, 50603 Kuala Lumpur, Malaysia |

**Abstract**

Room temperature ionic liquids show potential as an alternative to conventional organic membrane solvents mainly due to their properties of low vapor pressure, low volatility and they are often stable. In the present work, the technical feasibilies of room temperature ionic liquids as bulk liquid membranes for phenol removal were investigated experimentally. Three ionic liquids with high hydrophobicity were used and their phenol removal efficiency, membrane stability and membrane loss were studied. Besides that, the effects of several parameters, namely feed phase pH, feed concentration, NaOH concentration and stirring speeds on the performance of best ionic liquid membrane were also evaluated. Lastly, an optimization study on bulk ionic liquid membrane was conducted and the maximum phenol removal efficiency was compared with the organic liquid membranes. The preliminary study shows that high phenol extraction and stripping efficiencies of 96.21% and 98.10%, respectively can be achieved by ionic liquid membrane with a low membrane loss which offers a better choice to organic membrane solvents.

**Keywords**

Phenol, Ionic liquid, Liquid membrane, Extraction efficiency, Stripping efficiency

## 1. Introduction

Phenol is an organic substance which is used in several manufacturing processes such as for the manufacture of phenolic resin and other phenol derivative chemicals. It is also used as a solvent, as an antiseptic and as additive in disinfectant [1]. According to the Health and Safety Guide No88 [2], phenol is toxic and has a carcinogenic effect, therefore of considerable health concern to human and aquatic lives when released to the environment. The concentrations of phenol in industrial effluents are normally in a range of 2.8-6800ppm [1,3] which is much higher than the $LD_{50}$ for aquatic lives [2]. This gives adverse effect and disrupts the aquatic ecosystem balance. Department of Environment Malaysia has set a maximum phenol discharge concentration of 1ppm [4]. Thus, there is an urgent need to find an effective method for the treatment of wastewater containing phenol.

Among the available treatment methods, liquid membrane also attracts attention from the industries. Liquid membrane is defined as the liquid that serves as semi-permeable liquid which forms a barrier between the feed phase and the stripping phase and it transports the targeted solute selectively from one phase to another [5, 6]. This method combines the extraction and stripping processes in a single stage, thus providing lower capital and operating cost, technical simplicity, and independency on the transport equilibrium limitation [7-11]. Applications of liquid membranes in separation field have been reported by a number of researchers, mainly for the separation of organic compounds [3, 6, 8, 12] and metal ions [9, 13]. However, the utilization of organic membrane solvents such as hydrocarbons and chloroalkanes [3, 6, 8-10, 12-13] have their drawbacks as they are usually volatile, flammable and often toxic, thus restricting their applications in the industries.

Room temperature ionic liquids are a group of low melting point salts that consist of organic cations and organic/inorganic anions. They have negligible vapor pressure, low flammability and they can be fine tuned according to the design purpose. These unique properties serve them as an option to replace the volatile organic solvents. In this study, imidazolium based ionic liquids were chosen as the data on their physical and chemical properties are more readily available in comparison to other ionic liquids [14-17]. In fact, studies on liquid-liquid extraction of phenol by imidazolium ionic liquids with $[BF_4^-]$ and $[PF_6^-]$ anions have shown that they are better than the organic solvents in terms of extraction efficiency [3, 18-21]. The works of Li et al. [18], Vidal et al. [19] and Fan et al. [20] have proven that the combination of cations and anions strongly affect the extraction efficiency. In addition, the effects of several parameters such as pH, operating temperature and volume ratio of ionic liquid/water on the extraction efficiency have also been evaluated [18-21].

Despite the favorable phenol extraction efficiency in liquid-liquid extraction, the utilization of imidazolium based ionic liquid membrane for phenol removal is less studied. Instead, they are some published information on the separation of other organic compounds such as toluene [22], aromatic hydrocarbons [23], trans-esterification products [11, 24] and selective separation of 1-butanol, 1-propanol, cyclohexanol, 1,4-dioxane, cyclohexanone, morpholine and methylmorpholine [25] by these ionic liquids. These works show that imidazolium based ionic liquid membranes provide selective transport with good stability. However, Fortunato et al. [26-27] and Matsumoto et al. [28] reported that the selectivities of ionic liquid membrane diminished when contacted with aqueous phase. According to them, this was mainly caused by the significant water solubility of these ionic liquids. The formation of water microenvironment path in imidazolium based ionic liquids with [$BF_4^-$] and [$PF_6^-$] anions reduced the membrane selectivities during the transport of water soluble compounds such as tritiated water, NaCl, amino acid, amino acid esters [26-27] and penicillin G [28]. However, Fortunato et al. [26] also reported that the selectivity was not diminished entirely because the transport of thymol blue was less affected by the presence of water microenvironment.

In the present work, the technical feasibility of hydrophobic ionic liquids 1-butyl-3-methylimidazolium hexafluorophosphate [Bmim][$PF_6$], 1-butyl-3-methylimidazolium bis(trifluoromethylsulfonyl)imide [Bmim][$NTf_2$] and 1-butyl-3-methylimidazolium tris(pentafluoroethyl)trifluorophosphate [Bmim][FAP] when applied as bulk ionic liquid membranes for phenol removal was discussed. Bulk liquid membrane system was chosen as its simple configuration enabled the observation on the effect of ionic liquids on the phenol extraction and stripping efficiencies. The effect of different types of anion and hydrophobicity of the ionic liquids on the phenol extraction and stripping efficiencies were investigated. From there, the best ionic liquid which provided the highest performance was chosen and its behavior as liquid membrane under different operating parameters of feed phase pH, feed concentration, NaOH concentration and stirring speeds was studied. Furthermore, an optimization study of ionic liquid membrane for maximum overall phenol recovery was carried out and its performance was compared with kerosene and dichloromethane, the organic liquid membranes.

## 2. Experimental

*2.1 Materials*

The phenol crystals and ionic liquids 1-butyl-3-methylimidazolium hexafluorophosphate [Bmim][$PF_6$], 1-butyl-3-methylimidazolium bis(trifluoromethylsulfonyl)imide [Bmim][$NTf_2$] and 1-butyl-3-methylimidazolium tris(pentafluoroethyl)trifluorophosphate [Bmim][FAP] were supplied by Merck Sdn Bhd. NaOH pellets, HCl and dichloromethane were supplied by R&M Chemicals while kerosene was supplied by ACROS Organics. Phenol absorbance measurement was accomplished by using UV-Vis Spectrophotometer SECOMAM UviLine 9400 while IKA Lab Egg Stirrer RW11 and magnetic stirrer MSH-20D were used as parts of bulk liquid membrane systems. pH meter Cyber Scan pH300 and Rheintacho Rotaro Tachometer were also used in the experiments, mainly for pH measurement and stirring speed determination, respectively.

*2.2 Preparation of solutions*

Phenol solutions with different concentrations were prepared in a calibrated volumetric flask by dissolving the theoretical amount of phenol crystals into distilled water. Each was stirred for approximately 10 minutes to ensure that the phenol dissolved completely. The same procedure was applied for the preparation of NaOH solution from the pellets. HCl solution was prepared by diluting the concentrated HCl with distilled water in a volumetric flask. The solution was stirred to ensure a well mixed solution.

*2.3 Pre-treatment of ionic liquids*

Initially, ionic liquids, [Bmim][$NTf_2$], [Bmim][$PF_6$] and [Bmim][FAP], used as membrane solvents were washed with distilled water to remove any trace contaminants. The ionic liquids were then heated in a dryer at 75$^o$C for 2 days to remove any excess moisture [20] and then were stored in a desiccator before usage. [Bmim][$PF_6$] does not undergo heating process as wet [Bmim][$PF_6$] was not stable under high temperature condition.

*2.4 Bulk ionic liquid membrane experiment*

The experiment was carried out in a rectangular glass cell with an inner dimension of 12.1cm length, 5.6cm width and 11.9cm height. The cell was separated into two equal volume compartments by a partition of about 0.2cm thickness in the middle of the cell with a 0.80 cm opening at the bottom to enable transport of phenol from one compartment to another. A volume of 80mL of ionic liquid was weighed and transferred into the cell above the bottom clearance. After that, 200mL each of 300ppm phenol solution and 0.5M NaOH solution was transferred into the feed compartment and stripping compartment, respectively, as shown in Fig. 1 where both phases were bridged by the ionic liquid. The aqueous phases were stirred by IKA Lab Egg Stirrers RW11 with paddle blade of 3.4cm diameter at 200rpm while the membrane phase was stirred by a magnetic stirrer MSH-20D with a magnetic bar of 2.5cm diameter at a fixed speed of 100rpm. The duration of each experiment was 5 hours. At different time interval, 1mL of sample was taken from the feed phase and stripping phase using a calibrated pipette. The samples were then analyzed by UV-Vis Spectrophotometer SECOMAM UviLine 9400, and both extraction and stripping efficiencies were calculated, as discussed in Section 2.5. The experiments were at least duplicated and the results obtained were within an experimental error/deviation range of 4%.

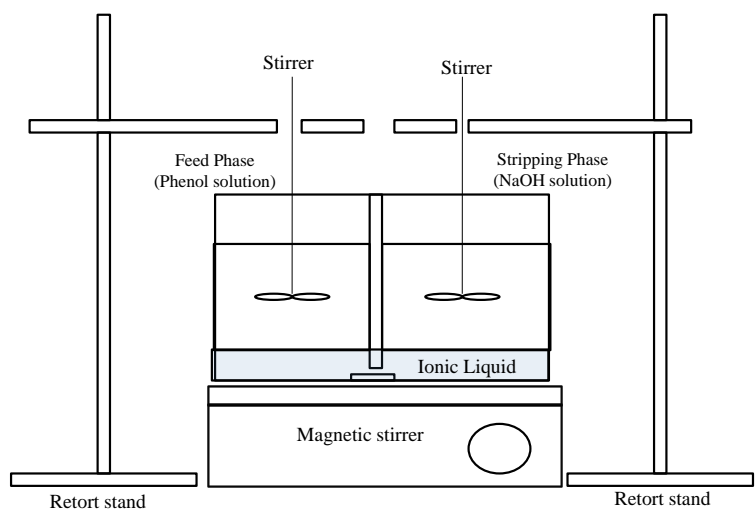

Fig. 1. Experiment setup for bulk ionic liquid membrane

*2.5 Analytical method*

*2.5.1 Dilution of samples*

The samples were diluted before analysis to make sure the absorbances of the samples can be measured within detectable range of the UV-Vis Spectrophotometer. The feed samples were diluted with 0.5M HCl solution which provided an acidic environment, whereby only molecular phenol existed while the stripping samples were diluted with 0.5M NaOH solution, whereby phenol only existed in the form of phenolate ion (sodium phenolate). Samples' absorbances were measured in wavelength of 270nm for molecular phenol detection in the feed samples and 288nm for phenolate ion detection in the stripping samples [29-30]. The concentrations of phenol and phenolate in the samples were obtained by referring to their linear absorbance-concentration calibration curves, ranging from 0 to 300ppm and 0 to 185ppm, respectively.

*2.5.2 Determination of extraction and stripping efficiencies in bulk ionic liquid membrane*

Phenol extraction and stripping efficiencies were determined by the concentration of phenol in each phase. The extraction efficiency was calculated using Equation (1):

$$\text{Extraction efficiency, \%} = \frac{\text{Initial concentration - Concentration of feed samples}}{\text{Initial concentration}} \times 100 \quad (1)$$

The stripping efficiency was calculated using Equation (2):

$$\text{Stripping efficiency, \%} = \frac{\text{Phenol concentration in stripping samples}}{\text{Initial concentration - Concentration of feed samples}} \times 100 \quad (2)$$

The amount of phenol in the stripping phase was determined by converting the amount of sodium phenolate that was detected in the stripping phase to the theoretical amount of phenol that should be present in the stripping phase, based on the mole balance from Equation(3).

$$C_6H_5OH + NaOH \rightarrow C_6H_5ONa + H_2O \quad (3)$$

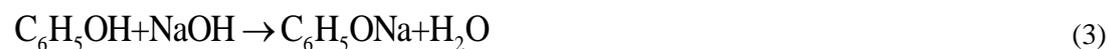

*2.6 Reuse of ionic liquids*

Ionic liquid was regenerated according to the method that was published by Fan et al. [20]. The ionic liquid was mixed vigorously for 30 minutes with 150mL of 0.5M NaOH solution, which acted as a stripping agent to remove phenol. After phase separation, the aqueous sample was analyzed for sodium phenolate concentration using the UV-Vis spectrophotometer. This procedure was repeated until there was no significant sodium phenolate was detected in the NaOH solution. The ionic liquid was then washed with distilled water until a pH of 6.5-7.5 was obtained. The excess moisture in the replenished ionic liquid was removed by heating at 75$^{o}$C for 2 days [20]. Ionic liquid was then stored in a desiccator before it was being reused. Again, it was worth to be noted that high temperature drying process was not conducted for [Bmim][PF$_6$] as the hydrolysis process was observed when wet [Bmim][PF$_6$] was heated. For experiment purpose, wet [Bmim][PF$_6$] was used.

## 3. Results and discussion

*3.1 Effect of different types of ionic liquids as membrane solvents*

The performance of bulk ionic liquid membrane by different types of ionic liquids was evaluated in terms of phenol extraction efficiency, stripping efficiency, membrane stability and membrane loss.

*3.1.1 Extraction efficiency*

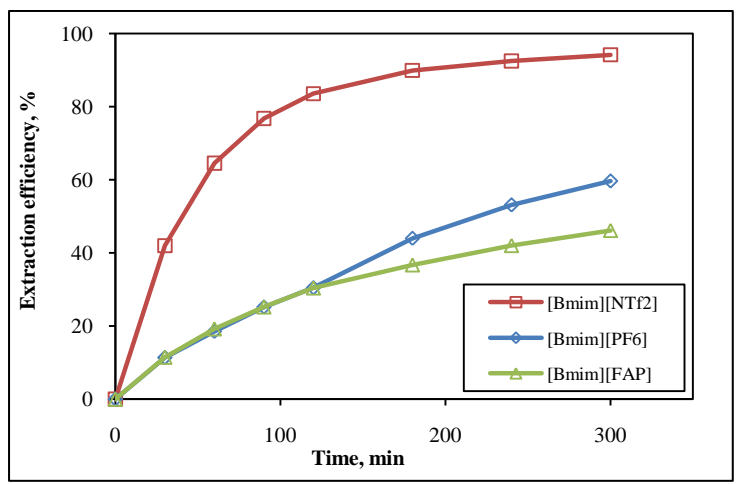

Fig. 2. Extraction efficiency of phenol by bulk ionic liquid membrane (Feed phase pH: ≈6.5; Feed concentration: 300ppm; NaOH concentration: 0.5M)

Ionic liquids [Bmim][PF$_6$], [Bmim][NTf$_2$] and [Bmim][FAP] were utilized as membrane solvents in this study. Fig. 2 shows that the extraction efficiency of bulk ionic liquid membrane is in the order of [Bmim][NTf$_2$] > [Bmim][PF$_6$] > [Bmim][FAP]. Although the work of Fan et al. [20] claimed that the increment of ionic liquids' hydrophobicity and hydrogen bond basicity strength increased phenol extraction efficiency, the present work showed that the hydrophobicity of ionic liquids had less effect on governing the phenol extraction efficiency. Among the ionic liquids studied, [Bmim][FAP] was the most hydrophobic ionic liquid [31-32] but it showed the poorest phenol extraction efficiency. The hydrophobicity of ionic liquid contributed by the anions was less significant on governing the phenol extraction process. In contrast, the extraction efficiency was found to be more dependent on the strength of the hydrogen bond basicity of the ionic liquids, in the sequence of [Bmim][NTf$_2$] > [Bmim][PF$_6$] > [Bmim][FAP] [15, 33-34].

*3.1.2 Stripping efficiency*

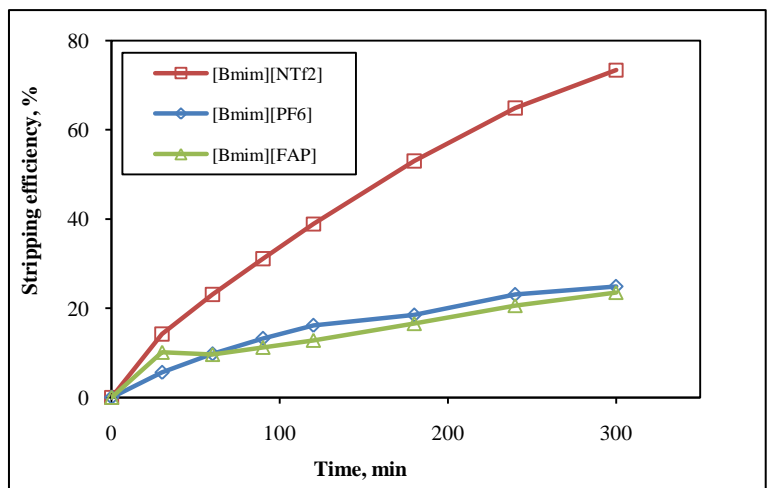

Fig. 3. Stripping efficiency of phenol by bulk ionic liquid membrane (Feed phase pH: ≈6.5; Feed concentration: 300ppm; NaOH concentration: 0.5M)

Fig. 3 shows that phenol in the ionic liquid membrane is stripped down by NaOH effectively. However, it was found that the stripping efficiencies were differed, whereby the efficiency of [Bmim][NTf$_2$] was greater than [Bmim][PF$_6$] which was the same as [Bmim][FAP]. In bulk liquid membrane process, the viscosity of the membrane solvent was significant on governing the stripping rate due to its high membrane thickness. From the published data [14-17, 31], the viscosity of ionic liquids studied are in the order of [Bmim][NTf$_2$] < [Bmim][FAP] < [Bmim][PF$_6$]. This was true for [Bmim][NTf$_2$] which had the highest stripping rate and efficiency among the ionic liquids studied as its viscosity was the lowest. However, it was found that ionic liquid membranes of [Bmim][PF$_6$] and [Bmim][FAP] had similar phenol stripping efficiency even though the viscosity of [Bmim][FAP] was much lower than [Bmim][PF$_6$]. This may be due to high hydrophobic property of [Bmim][FAP] which contributes to higher boundary layer thickness between the membrane phase and the aqueous stripping phase, resultant in a lower stripping efficiency. However, the actual reason for this observation remains unclear.

*3.1.3 Membrane stability*

The stabilities of ionic liquid membranes were examined through the reusability of regenerated ionic liquids and the transport of water soluble compound (NaOH) from stripping phase to the feed phase through the water microenvironment path within the ionic liquid.

Ionic liquid was regenerated using the method mentioned in Section 2.6 and then reused. It was found that the reused ionic liquid gave similar extraction and stripping efficiencies within the range of experimental error of 4%. This further indicated the possible change of structure of ionic liquids had less effect on the extraction and stripping efficiencies. In additions, the study also showed that the transport of NaOH to the feed phase by the water microenvironment path was not observed for all types of ionic liquids tested as the change of pH in the feed phase during the experiments was insignificant. Unlike the work of Fortunato et al. [26-27] and Matsumoto et al. [28], the transport of NaOH was not observed even for [Bmim][PF$_6$] based liquid membrane. This may be attributed to the fast experimental duration and phenol transport rate, where the formation of water microenvironment path was not fully developed. Besides that, the utilization of membrane stirring may be another factor that prevented the formation of the water microenvironment path, where the path was normally disrupted by the turbulence of membrane stirring.

*3.1.4 Membrane loss*

The loss of ionic liquid membrane was investigated and the percentage of ionic liquid loss is as shown in Table 1. It was found that the loss of ionic liquids was in an ascending order of [Bmim][FAP] < [Bmim][NTf$_2$] < [Bmim][PF$_6$]. As shown in Table 1, the loss of ionic liquid during the experiments is strongly dependent on the dissolution process of ionic liquid into the adjacent aqueous phases, where the higher the solubility of the ionic liquid in water, the higher the loss. The loss of ionic liquid membrane could be decreased by utilizing a more hydrophobic ionic liquid. However, a good phenol transport rate was not guaranteed.

Table 1: Percentage of ionic liquid loss

| Ionic liquid | Solubility in water, g/g | Ionic liquid loss, % |
|---|---|---|
| [Bmim][PF$_6$] | 0.020 [15] | 12.71% |
| [Bmim][NTf$_2$] | 0.0072 [15] | 7.41% |
| [Bmim][FAP] | <0.0072 [31-32] | 4.74% |

*3.2 Behavior of ionic liquid as bulk liquid membranes for phenol removal*

The results in Section 3.1 show that ionic liquid membrane based on [Bmim][NTf$_2$] gives higher extraction and stripping efficiencies than [Bmim][PF$_6$] and [Bmim][FAP]. Hence, [Bmim][NTf$_2$] was further studied on the effect of feed phase pH, feed concentration, NaOH concentration and stirring speeds on the performance of ionic liquid membrane.

*3.2.1 Effect of feed phase pH*

The feed phase pH was adjusted using HCl and NaOH solutions before the experiment. The form of phenol that was existed in the feed phase was strongly dependent on the feed phase pH. Majority forms of phenol in this study under different feed phase pH are as shown in Equations (4) and (5).

At pH < 9.23 (< pKa): 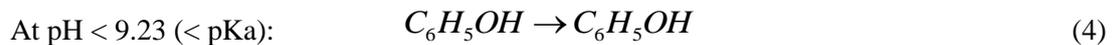 $\quad C_6H_5OH \rightarrow C_6H_5OH$ \hfill (4)

At pH > 9.23 (> pKa): 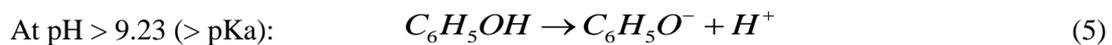 $\quad C_6H_5OH \rightarrow C_6H_5O^- + H^+$ \hfill (5)

Fig. 4 shows that the extraction efficiency of phenol by [Bmim][NTf$_2$] based liquid membrane remains constant at 94% when the feed phase pH is held below 6.56. However, there was a drastic decrease in the extraction efficiency when the pH was increased from 6.56 to 9.23 and 11.19. This was also observed by Vidal et al. [19], Fan et al. [20] and Khachatryan et al. [21] in liquid-liquid extraction process. According to Fan et al. [20], the extraction of phenol was dependent on the formation of hydrogen bonding between the molecular phenol with the anion of the ionic liquid. The hydrogen bonding was weakened as the feed phase pH was increased where phenol, being a weak acid, ionized as phenolate ions ($C_6H_5O^-$) under high pH condition ($\geq$ pKa value of 9.89 [1]).

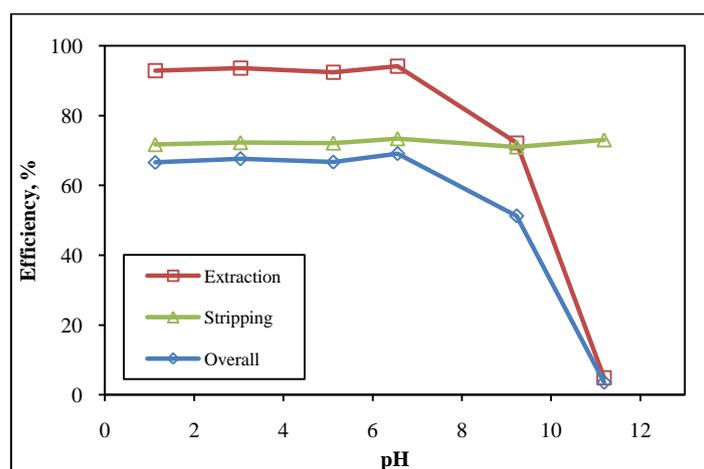

Fig. 4. Effect of feed phase pH on the performance of bulk ionic liquid membrane (Feed concentration: 300ppm; NaOH concentration: 0.5M)

Unlike the mechanism that was suggested by Khachatryan et al [21] for [Bmim][PF$_6$], the anion exchange between [NTf$_2^-$] and phenolate ion was not observed in high pH of 11.19 as a steep reduction in the extraction efficiency was observed. This was attributed to the unfavorability on the formation of hydrogen bonding between the phenolate ion with [Bmim][NTf$_2$]. Besides that, Fig. 4 shows that the stripping efficiency is less affected by the feed phase pH, whereby the efficiency is recorded in the range of 71-73% for pH of 1.13-11.19. Unlike other ionic liquids like tetrahexylammonium dihexylsulfosuccinate and trioctylmethylammonium salycylate which extracted phenol in both molecular and anion forms [35], the low extraction efficiency for phenolate ion by [Bmim][NTf$_2$] ensured a stable one way transport process was achieved in high pH stripping phase.

*3.2.2 Effect of feed concentration*

Fig. 5 illustrates the feed concentration has less effect on the extraction efficiency. The final extraction efficiency remained 94% as the feed concentration was increased from 150-6000ppm. This may be caused by the usage of high amount of ionic liquid in bulk liquid membrane system which gave a high phenol dissolve capacity for phenol transport as a high upper limit for phenol saturation was contributed by the ionic liquid membrane. In additions, the simultaneous stripping process in liquid membrane system also prolonged the time required for membrane saturation. Thus, the extraction efficiency was not disturbed. However, a different trend was observed for the stripping efficiency where an obvious increment in the stripping efficiency was found as the feed concentration was increased from 2000-6000ppm, as shown in Fig. 6. Similar observation was also reported by Nabieyan et al. [10] for the passive transport of iodine in bulk liquid membrane.

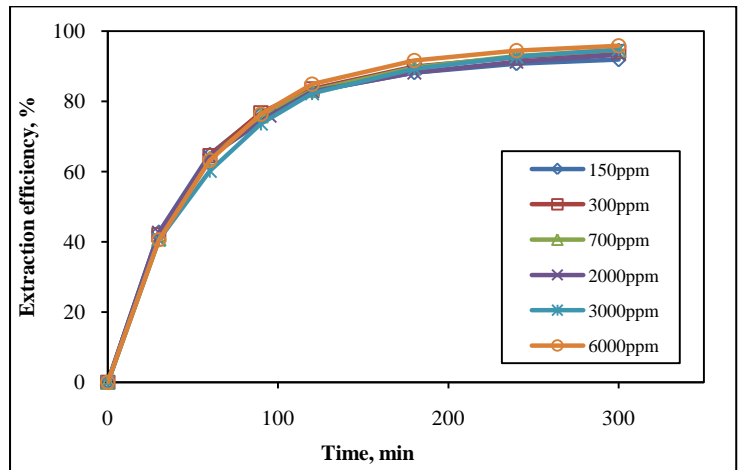

Fig. 5. Extraction efficiency of phenol by bulk ionic liquid membrane at different feed concentrations (Feed phase pH: ≈5.8-6.5; NaOH concentration: 0.5M)

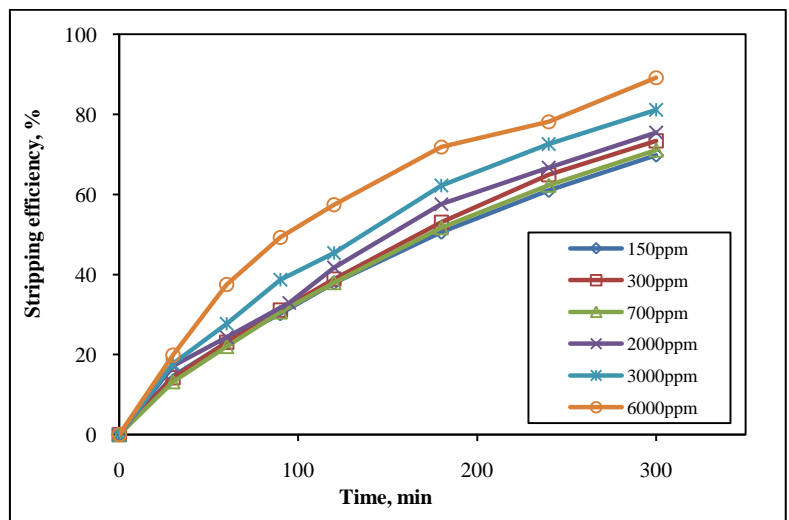

Fig. 6. Stripping efficiency of phenol by bulk ionic liquid membrane at different feed concentrations (Feed phase pH: ≈5.8-6.5; NaOH concentration: 0.5M)

An explanation on the stripping behaviour can be proposed with the aid of physical observation on the change of ionic liquid's tone during the transport of phenol in a set of experiments. Fig. 7 shows the diffusion process of phenol in bulk ionic liquid membrane at different feed concentration levels. At low phenol feed concentration (150-700ppm), the stripping rate was fast enough to prevent the built-up of phenol concentration in the membrane/stripping interface due to low amount of phenol was available in the membrane phase. Hence, the phenol that was distributed to the membrane/stripping interface utilized only a small amount of contact area for the stripping process. On the other hand, higher extraction rate was observed in high feed concentration as there was a high capacity for the phenol extraction in the ionic liquid membrane phase. The uneven high extraction and low stripping rates caused the built-up of phenol concentration in the membrane phase. With the aid of magnetic stirring in the membrane phase, significant amount of phenol accumulated in the membrane phase was distributed to a further distance in the stripping compartment, as shown in Fig. 7. This eventually increased the effective contact area between membrane/stripping interface. Thus, the stripping rate and efficiency were increased as the feed concentration was increased from 2000-6000ppm.

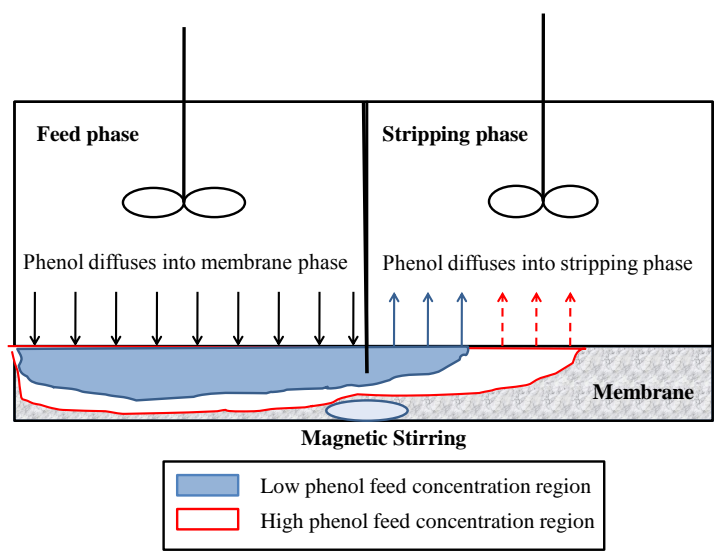

Fig. 7. Diffusion of phenol in bulk ionic liquid membrane at different feed concentrations

*3.2.3 Effect of NaOH concentration*

Fig. 8 illustrates the extraction efficiency of bulk ionic liquid membrane is less affected by the concentration of NaOH as long as it is available. A similar trend on the increment of the extraction efficiency with time was observed and a final extraction efficiency of 94% was achieved for the NaOH concentration range of 0.005-0.5M while they had significant influence on the stripping efficiency of the system. Fig. 9 shows that the stripping efficiency increases in the following order: 0M < 0.005M < 0.01M < 0.05M. In this study, NaOH concentration of 0.05M (pH ≈ 12.7) was observed to be adequate to maintain the high stripping efficiency and beyond this concentration, the stripping efficiency remained the same. However, the experiments for 0.005M and 0.01M (both pH < 12.2) showed a lower final stripping efficiency, and this could well be due to the insufficiency of NaOH in the stripping phase. Theoretically, after the diffusion process, phenol reacts with NaOH in the stripping phase to form sodium phenolate, and the activity of molecular phenol in the stripping phase is suppressed [1]. In other words, the activity of unreacted molecular phenol slowed down the stripping process by reducing the concentration gradient between membrane and stripping phases under low NaOH concentration, thus reducing the stripping rate and efficiency. This is supported by the works of Le *et al*. [8] and Xiao *et al*. [29] who claimed that the stripping process can be optimized and the incomplete reaction can only be prevented by maintaining pH > pKa by at least 2 units (high NaOH concentration) for the stripping phase.

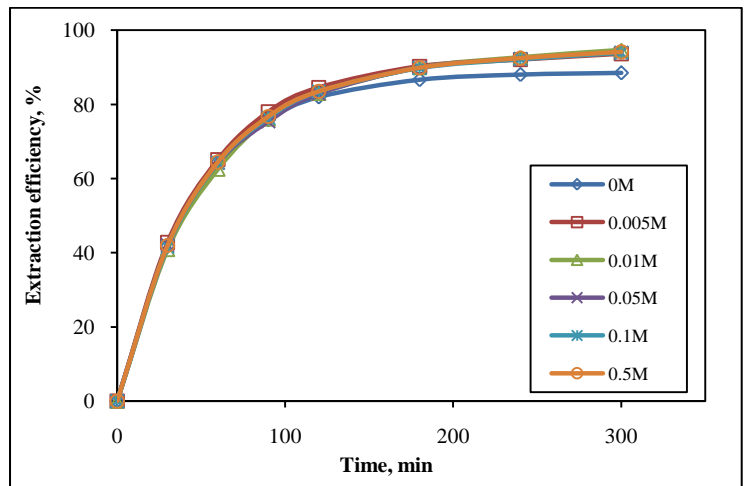

Fig. 8. Extraction efficiency of phenol by bulk ionic liquid membrane at different NaOH concentrations (Feed phase pH: ≈6.5; Feed concentration: 300ppm)

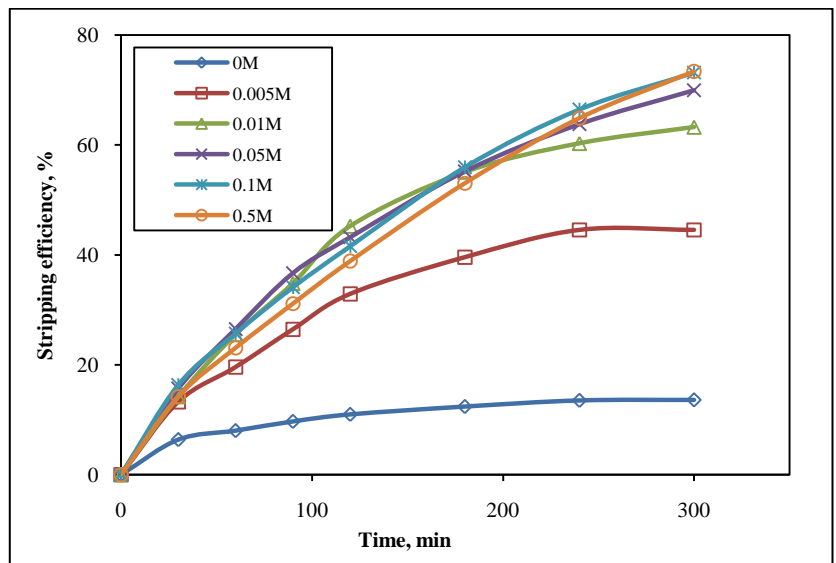

Fig. 9. Stripping efficiency of phenol by bulk ionic liquid membrane at different NaOH concentrations (Feed phase pH: ≈6.5; Feed concentration: 300ppm)

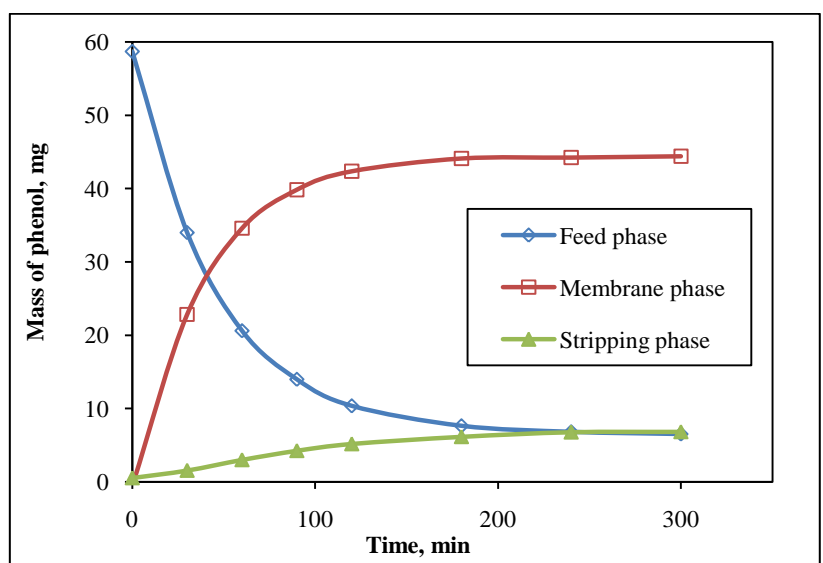

Fig. 10. Mass of phenol in each phase versus time for 0M NaOH concentration (Feed phase pH: ≈6.5; Feed concentration: 300ppm)

On the other hand, Figs. 8 and 9 show a lower performance for 0M NaOH concentration with a final extraction efficiency of 88.49% and stripping efficiency of 13.65%. A further study on the distribution of phenol in all three phases with time was conducted as illustrated in Fig. 10. The amount of phenol in the feed phase and the stripping phase eventually reach the same value of about 6.7mg and phenol saturation in membrane phase was observed at 44g. In other words, 77% of phenol was present in the membrane phase and is in line with the result for liquid-liquid extraction of phenol by [Bmim][NTf$_2$] (Unpublished work). This further justifies that the extraction and stripping processes in this experiment were dependent on the distribution coefficient of phenol between [Bmim][NTf$_2$] and both aqueous phases (distilled water). In this transport process, when phenol in the membrane phase reached the membrane/stripping interface, it diffuses into the stripping phase (distilled water) based on the difference in the concentration gradient of phenol between the phases. Since the activity of phenol was not suppressed by NaOH, the molecular phenol in the stripping phase started to reduce the stripping rate via reduction on the concentration gradient between the membrane phase and the stripping phase. This process continued until the amount of phenol in the stripping phase reached an extraction equilibrium with [Bmim][NTf$_2$] and the stripping rate became zero. As the characteristic of the stripping phase for 0M NaOH was the same as the feed phase (distilled water), a similar amount of phenol was detected in both of the feed phase and the stripping phase at the end of the experiment.

*3.2.4 Effect of stirring*

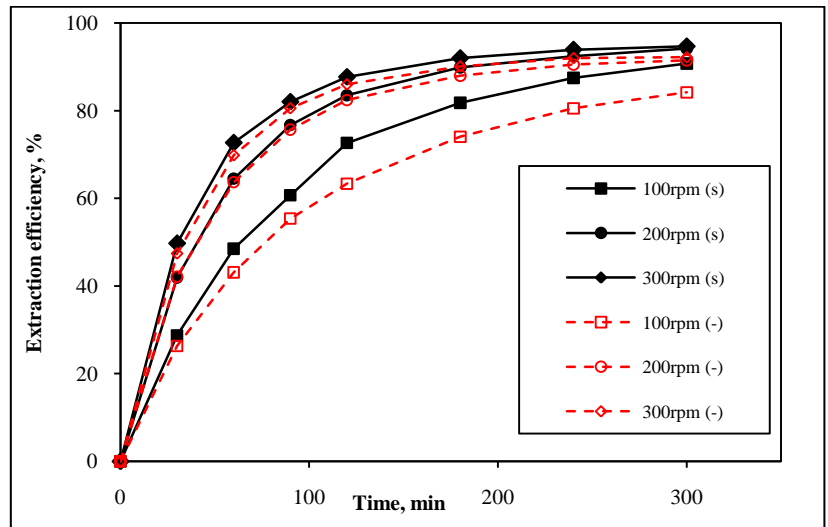

Fig. 11. Effect of stirring speed on the extraction efficiency of phenol by bulk ionic liquid membrane, (s): membrane stirring, (-): without membrane stirring (Feed phase pH: ≈6.5; Feed concentration: 300ppm; NaOH concentration: 0.5M)

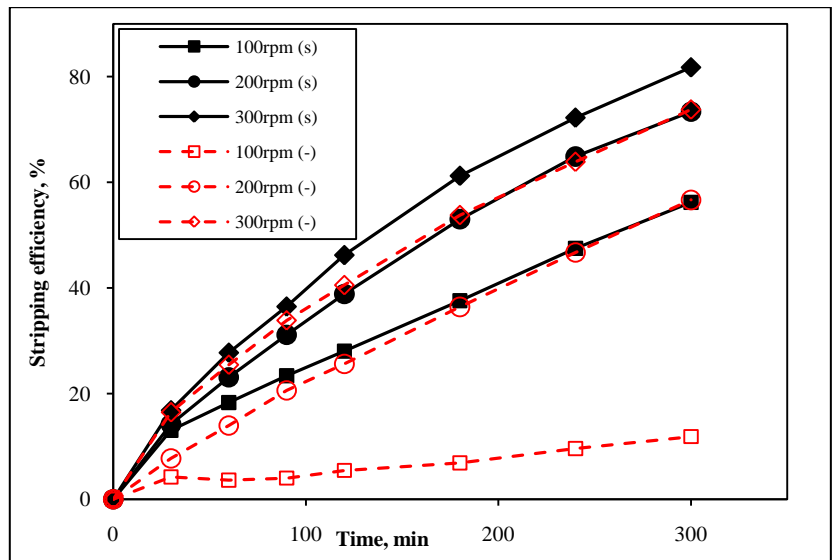

Fig. 12. Effect of stirring speed on the stripping efficiency of phenol by bulk ionic liquid membrane, (s): membrane stirring, (-): without membrane stirring (Feed phase pH: ≈6.5; Feed concentration: 300ppm; NaOH concentration: 0.5M)

Fig. 11 shows that the increment of aqueous stirring speed from 100-300rpm increases the rate of extraction efficiency. This was observed in both experiments with membrane stirring and without membrane stirring. Higher aqueous stirring speed increased the rate of extraction efficiency by providing a better mixing and minimized the boundary layer thickness between the membrane phase and aqueous phase [9]. Further, the presence of membrane stirring boosted the rate of extraction efficiency under the same aqueous stirring speed, as illustrated in Fig. 11. However, this enhancement became lower as the aqueous stirring speed was increased. In comparison to the extraction efficiency, the rate of stripping efficiency was increased at a higher magnitude with an increment in the aqueous stirring speed from 100-300rpm, as observed in Fig. 12. Besides that, under the same aqueous stirring speed, the presence of membrane stirring significantly enhanced the rate of stripping efficiency. In this case, phenol was transported through the membrane phase at a faster rate in the presence of membrane stirring where the additional force in membrane phase compensated the high viscosity membrane and long diffusion path for bulk liquid membrane. However, this enhancement was reduced again as the aqueous stirring speed was increased.

Generally, the increment on the aqueous stirring speed increases the rates of extraction and stripping efficiencies. However, membrane entrainment was observed when the aqueous stirring speed was increased up to 300rpm. Similar problem had been reported by other researchers for organic liquid membranes [6, 22] and this further caused other undesirable effects such as the requirement of another post-treatment. In order to solve this issue, it is worth noting that similar rate of stripping efficiency can be achieved under lower aqueous stirring speed in the presence of membrane stirring. As shown in Fig. 12, the experiments for 200rpm (s) with 300 (-), and 100 (s) with 200 (-) give the identical stripping efficiency of 73% and 57%, respectively after 5 hours experiment. Hence, the use of high aqueous stirring speed can be prevented by fine tuning the aqueous stirring speed and the membrane stirring speed. This will be further discussed in Section 3.3.

*3.2.5 Effect of different parameters on the behavior of bulk ionic liquid membranes*

Bulk liquid membrane based on [Bmim][NTf$_2$] follows similar behavior to organic liquid membranes. Unlike organic solvents, ionic liquid [Bmim][NTf$_2$] consists of cation and anion, and they are bonded by ionic interaction strength. However, this nature in bonding did not affect its applicability as liquid membrane for phenol removal. The extraction of phenol was mainly governed by hydrogen bonding rather than ion-exchange mechanism, as discussed in Section 3.2.1. Besides, phenolate ion was not soluble in [Bmim][NTf$_2$] via ionic bonding even though both of them were in ionic forms. Thus, stable extraction and stripping processes were observed.

Simple mass transfer mechanism, similar to the organic liquid membranes was observed in the transport of phenol by [Bmim][NTf$_2$] based liquid membrane. Most of the observations in the experiments can be explained according to extraction equilibrium and Fick's law of diffusion elaborated in Sections 3.2.2 and 3.2.3. However, the high relative viscosity of [Bmim][NTf$_2$] often gave a low phenol stripping efficiency compared to the extraction efficiency. The mass transfer resistance provided by [Bmim][NTf$_2$] in bulk liquid membrane process is significant, and it serves as a major disadvantage on the application of ionic liquid as bulk liquid membrane. Even though the study on stirring speeds shows that stripping efficiency can be enhanced by utilizing higher speed, other problem such as membrane entrainment is in risk. Hence, further optimization studies on the stirring speed as detailed in Section 3.3 have been carried out.

*3.3 Optimization on bulk ionic liquid membrane*

A study on the interrelationship between the aqueous stirring speed and membrane stirring speed was conducted in order to acquire the best combination that provides the highest phenol extraction and stripping efficiencies under non-entrainment condition. Stirring speeds were varied between 0-300rpm while other operating parameters were held constant. The constant values are as shown in Table 2.

Table 2: Operating parameters for bulk ionic liquid membrane based on [Bmim][NTf$_2$]

| Parameters | Value |
|---|---|
| pH | ≈6.5 |
| Feed concentration | 300ppm |
| NaOH concentration | 0.5M |
| Temperature | 25°C |
| Experiment duration | 300 minutes |

The study was conducted statistically by using Central Composite Design based Response Surface Methodology which was provided in the Design Expert 6.0.5 software. The mathematical model for the overall phenol recovery, i.e. products of extraction efficiency and stripping efficiency was obtained based on the Analysis of Variance (ANOVA). The statistical criteria such as insignificant lack of fit, in range of predicted $R^2$ and adjusted $R^2$, and the diagnosis of residuals were fulfilled.

Based on the statistical criteria, the interrelationship of aqueous stirring speed and membrane stirring speed on governing the overall phenol recovery is as shown in Fig. 13. In general, the increment on both stirring speeds increase the overall phenol recovery in a reduced cubic trend. For the individual effect, the increment of aqueous stirring speed from 0rpm (Level -1) to 300rpm (Level 1) increased the overall phenol recovery. On the other hand, the increment of membrane stirring speed also increased overall phenol recovery, but in a higher magnitude, with the same effect as discussed in Section 3.2.4. However, it is worth noting that further increase of membrane stirring speed beyond 225rpm (Level 0.5) reduced the overall phenol recovery. Beyond this level, the high turbulence in the membrane phase created high vortex and finally mixed the feed phase and the stripping phase. Thus, the phenol transport process failed and overall phenol recovery was reduced.

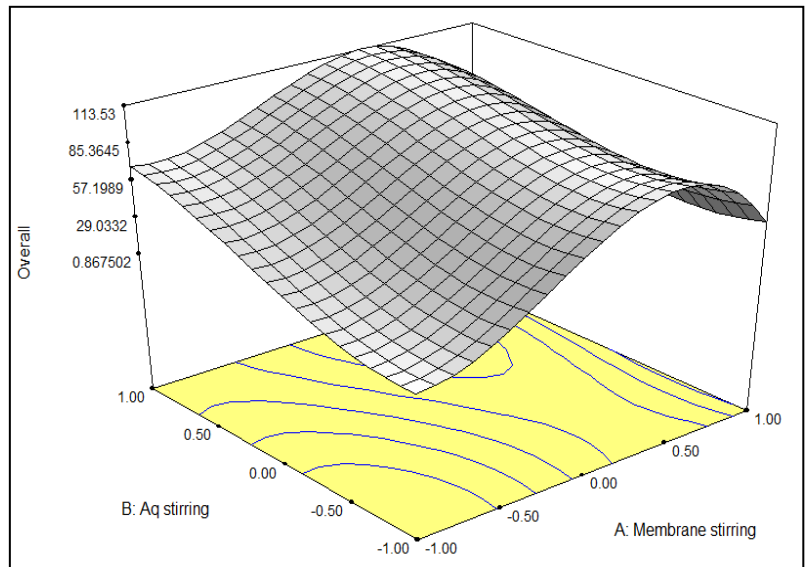

Fig. 13. 3D contour plot for overall phenol recovery

Table 3: Optimum stirring speeds for the best overall phenol recovery

| Parameters | Membrane stirring speed, rpm | Aqueous stirring speed, rpm |
|---|---|---|
| Optimum value (Coded) | -0.10 | 0.70 |
| Optimum value (Uncoded) | 135 | 255 |
| | **Predicted** | **Experimental** |
| Extraction efficiency, % | NA | 96.21 |
| Stripping efficiency, % | NA | 98.10 |
| Overall phenol recovery, % | 95.80 | 94.39 |

From the plot, the optimum stirring speeds that give the highest overall phenol recovery are in the range of 120-225rpm (Level -0.2 – 0.5) for membrane stirring and 150-300rpm (Level 0-1) for aqueous stirring. By the aid of Design Expert 6.0.5, a series of combinations on the stirring speeds that provide the best overall phenol recovery can be obtained from the plot. Take into account that the membrane entrainment was not detectable by the model, all of the combinations were justified experimentally. The best combination is as shown in Table 3.

Table 3 shows that the trend of interrelationship between both stirring speeds is valid as the predicted and experimental results were within an acceptable deviation of 1.41%. An overall phenol recovery as high as 94.39%, without membrane entrainment can be obtained experimentally using an aqueous stirring speed of 255rpm and membrane stirring speed of 135rpm.

*3.4 Efficiencies comparison with bulk organic liquid membranes*

The optimization of stirring speeds showed that a high phenol extraction and stripping efficiencies could be obtained without any membrane entrainment. In order to further justify the technical feasibility of [Bmim][NTf$_2$] as liquid membrane, an efficiency comparison was made between ionic liquid [Bmim][NTf$_2$] and organic membrane solvents. The extraction and stripping efficiencies of the liquid membranes are as shown in Fig. 14.

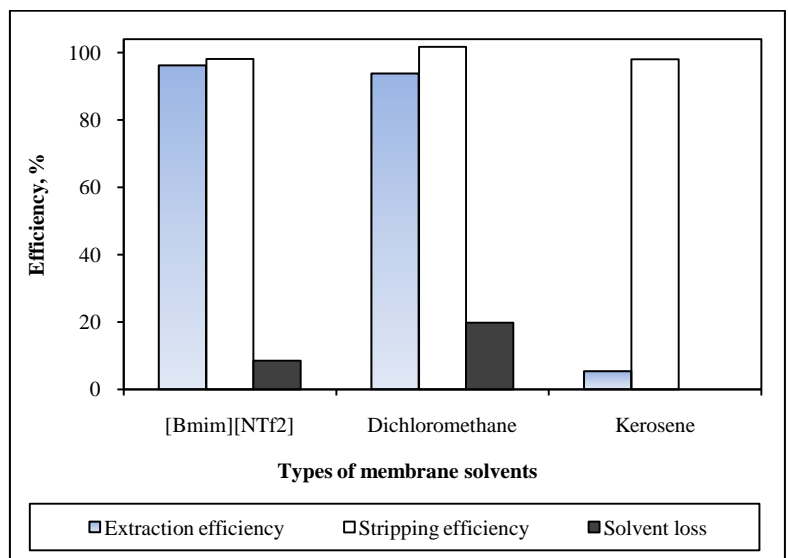

Fig. 14: Efficiency comparison for different types of membrane solvents in 5 hours experiment on their optimum stirring speeds (Feed phase pH: ≈6.5; Feed concentration: 300ppm; NaOH concentration: 0.5M)

Ionic liquid membrane based on [Bmim][NTf$_2$] provided a comparable extraction and stripping efficiencies with dichloromethane while it showed a better extraction efficiency as compared to kerosene at their optimum stirring speeds, respectively. It is worth noting that even though [Bmim][NTf$_2$] has a high relative viscosity in comparison to dichloromethane and kerosene in approximately two order magnitude, the effect of high viscosity can be reduced by fine-tuning the membrane stirring speed and the aqueous stirring speed. This is because [Bmim][NTf$_2$] is able to withstand a higher stirring speeds in comparison to organic membrane solvents. Consequently, there is a reduction in the mass transfer resistance and this facilitated [Bmim][NTf$_2$] to achieve comparable liquid membrane efficiencies.

Fig. 14 also shows that [Bmim][NTf$_2$] has lower solvent loss percentages of 8.48% in comparison to 19.76% of dichloromethane. This further explains that the characteristic of ionic liquid membrane which has low vapor pressure reduces solvent loss. From the foregoing, it is clear that ionic liquid is a better choice as an alternative to organic membrane solvents.

*3.5 Transport mechanism*

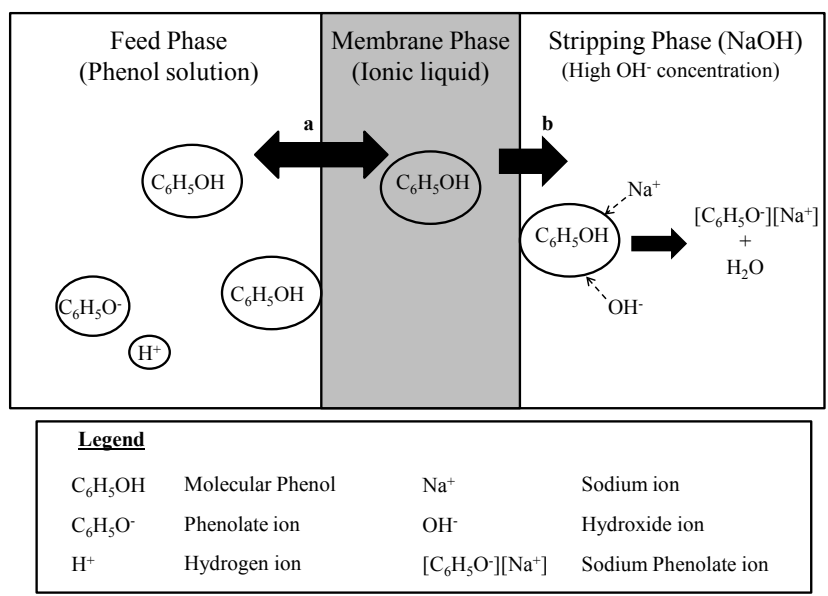

Fig.15: Transport mechanism of phenol in [Bmim][NTf$_2$] bulk ionic liquid membrane

Based on the results in Section 3.2, it is proposed that pertraction of phenol through ionic liquid membrane is occurred via simple permeation process. Fig. 15 illustrates the transport mechanism of phenol through [Bmim][NTf$_2$] bulk ionic liquid membrane. The study showed that only molecular phenol is transported through the ionic liquid membrane. Ionic form of phenol (phenolate ion) was remained in the feed phase as it was not extracted by ionic liquid membrane, as discussed in Section 3.2.1. General transport mechanism is as shown in Equation (6).

$$C_6H_5OH_{aq} \rightleftharpoons C_6H_5OH_{IL} \xrightarrow{NaOH} C_6H_5ONa + H_2O \qquad (6)$$

Pertraction of molecular phenol in [Bmim][NTf$_2$] bulk ionic liquid membrane consists of two major processes which are a) extraction and b) stripping, as labelled in Fig. 15:

a) Molecular phenol dissolves in ionic liquid membrane due to its high distribution coefficient between [Bmim][NTf$_2$] and water. Similar to liquid-liquid extraction, the extraction process is reversible which is governed by equilibrium extraction of phenol between the ionic liquid membrane and the feed phase [21].

b) Phenol is transported through the ionic liquid membrane to reach membrane/stripping interface. Molecular phenol diffuses into the stripping phase due to the concentration gradient difference between the phases. As a weak acid, phenol ionizes into phenolate ion and hydrogen ion under high OH$^-$ concentration stripping phase (high pH). These ions react with sodium ion and hydroxide ion in the stripping phase, respectively to yield sodium phenolate and water. As phenolate ion is insoluble in ionic liquid membrane, back stripping of phenol is prevented. Moreover, NaOH in the stripping phase suppresses phenol's activity in this phase and maintaining concentration gradient between membrane phase and stripping phase [1]. Thus, one way stripping process is achieved.

## 4. Conclusions

The results showed that ionic liquid [Bmim][NTf$_2$] which has high hydrogen bond basicity strength and low viscosity gave the best phenol extraction and stripping efficiencies compared to [Bmim][PF$_6$] and [Bmim][FAP]. The hydrophobicity of the ionic liquids did not have much effect on the results but contributed significantly in the membrane recovery whereby the increment in the hydrophobicity reduces the ionic liquid membrane loss. A good membrane stability and reusability of the ionic liquids were achieved for all the ionic liquids tested.

The effect of feed phase pH, feed concentration, NaOH concentration and stirring speeds showed that [Bmim][NTf$_2$] based liquid membrane behaved as the same manner as that of organic liquid membranes. A stable molecular phenol transport process was achieved by [Bmim][NTf$_2$] liquid membrane and this trend can be justified by the extraction equilibrium and Fick's law of diffusion.

The stripping efficiency lowering due to the high viscosity of [Bmim][NTf$_2$] was solved by regulating the aqueous stirring speed and the membrane stirring speed. A high extraction and stripping efficiencies of 96.21% and 98.10% were obtained in the absence of membrane entrainment at the optimum aqueous stirring speed of 255rpm and membrane stirring speed of 135rpm, respectively. In comparison to the organic membrane solvents, [Bmim][NTf$_2$] showed better performance to kerosene but comparable to dichloromethane.

Furthermore, the loss of [Bmim][NTf$_2$] during the experiments was much lower in comparison to the loss in dichloromethane. Hence, [Bmim][NTf$_2$] offers a better choice as it gives high phenol extraction and stripping efficiencies with a stable transport process and lower solvent loss.

**Acknowledgment**

This study has been supported by University of Malaya through the UMRG grant RG033/09SUS and PPP grant PS138/2009C.